\documentclass[preprint]{aastex63}

\accepted{2021 April 5}
\submitjournal{The Planetary Science Journal}

\shortauthors{Cantillo et al.}
\graphicspath{{./}{figures/}}

\begin{document}

\title{Constraining the Regolith Composition of Asteroid (16) Psyche via Laboratory Visible Near-Infrared Spectroscopy}

\correspondingauthor{David C. Cantillo}
\email{davidcantillo@email.arizona.edu}

\author{David C. Cantillo}
\affiliation{Lunar and Planetary Laboratory \\
University of Arizona \\
1629 E University Blvd \\
Tucson AZ 85721-0092, USA}

\author[0000-0002-7743-3491]{Vishnu Reddy}
\affiliation{Lunar and Planetary Laboratory \\
University of Arizona \\
1629 E University Blvd \\
Tucson AZ 85721-0092, USA}

\author{Benjamin N. L. Sharkey}
\affiliation{Lunar and Planetary Laboratory \\
University of Arizona \\
1629 E University Blvd \\
Tucson AZ 85721-0092, USA}

\author{Neil A. Pearson}
\affiliation{Planetary Science Institute \\
1700 E Fort Lowell Suite 106 \\
Tucson, AZ, 85719, USA}

\author{Juan A. Sanchez}
\affiliation{Planetary Science Institute \\
1700 E Fort Lowell Suite 106 \\
Tucson, AZ, 85719, USA}

\author[0000-0001-5456-2912]{Matthew R. M. Izawa}
\affiliation{Institute for Planetary Materials\\
Okayama University \\
827 Yamada, Misasa, Tottori 682-0193, Japan}

\author[0000-0003-1008-7499]{Theodore Kareta}
\affiliation{Lunar and Planetary Laboratory \\
University of Arizona \\
1629 E University Blvd \\
Tucson AZ 85721-0092, USA}

\author{Tanner S. Campbell}
\affiliation{Lunar and Planetary Laboratory \\
University of Arizona \\
1629 E University Blvd \\
Tucson AZ 85721-0092, USA}

\author{Om Chabra}
\affiliation{Catalina Foothills High School \\
4300 E Sunrise Dr \\
Tucson, AZ 85718, USA}

\begin{abstract}
(16) Psyche is the largest M-type asteroid in the main belt and the target of the NASA Discovery-class \textit{Psyche} mission. Despite gaining considerable interest in the scientific community, Psyche’s composition and formation remain unconstrained. Originally, Psyche was considered to be almost entirely composed of metal due to its high radar albedo and spectral similarities to iron meteorites. More recent telescopic observations suggest the additional presence of low-Fe pyroxene and exogenic carbonaceous chondrites on the asteroid’s surface. To better understand the abundances of these additional materials, we investigated visible near-infrared (0.35 - 2.5 $\mu m$) spectral properties of three-component laboratory mixtures of metal, low-Fe pyroxene, and carbonaceous chondrite. We compared the band depths and spectral slopes of these mixtures to the telescopic spectrum of (16) Psyche to constrain material abundances. We find that the best matching mixture to Psyche consists of 82.5\% metal, 7\% low-Fe pyroxene, and 10.5\% carbonaceous chondrite by weight, suggesting that the asteroid is less metallic than originally estimated ($\sim$94\%). The relatively high abundance of carbonaceous chondrite material estimated from our laboratory experiments implies the delivery of this exogenic material through low velocity collisions to Psyche’s surface. Assuming that Psyche’s surface is representative of its bulk material content, our results suggest a porosity of 35\% to match recent density estimates.

\end{abstract}

\section{Introduction} \label{sec:intro}
 
\begin{figure}[ht!]
\plotone{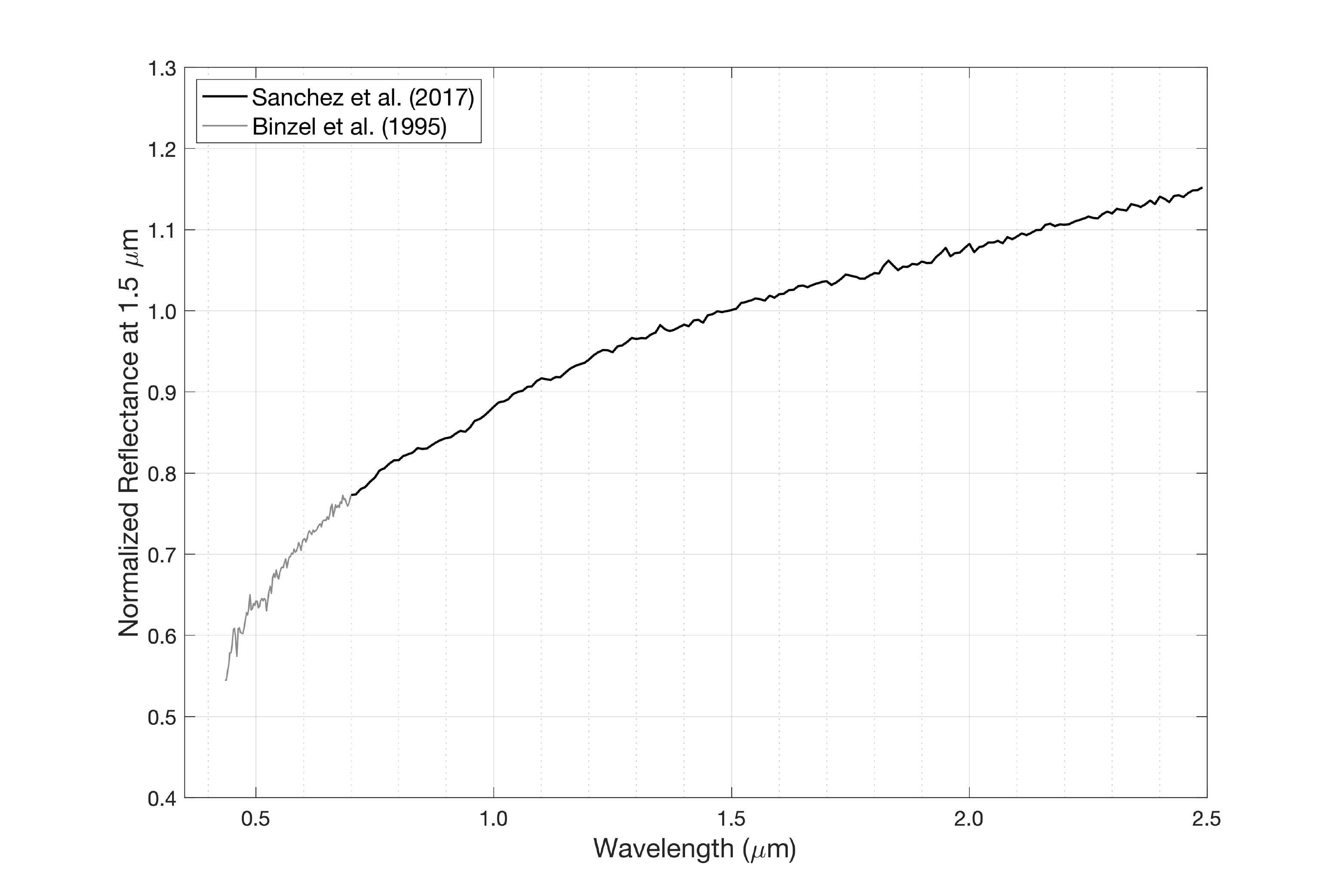}
\caption{Visible near-infrared telescopic spectrum of Main Belt asteroid (16) Psyche using ground-based telescopes. Data from 0.435 to 0.7 $\mu m$ (light gray) was taken with the McGraw Hill Observatory 2.4 m Hiltner telescope on Kitt Peak by \citet{Binzel1995}). Near-infrared data (black) was acquired from the NASA Infrared Telescope Facility (IRTF) by \citet{Sanchez2017}. The data are normalized to unity at 1.5 $\mu m$.}
\label{fig:psyche_vnir}
\end{figure}

Asteroids are time capsules that provide insight into fundamental questions regarding the creation of protoplanets and the processes that transformed them into the planetary bodies observed today. Many of the planetesimals in the main belt are thought to be derived from partially or fully differentiated bodies early in our solar system history. Some metal-rich (M-type) asteroids are unique as they may be derived from planetesimal cores \citep{Ostro1985}. Asteroid (16) Psyche is the largest M-type asteroid in the main belt \citep{Tholen1984}, with an average diameter of 223 $\pm$ 4 km \citep[][see Table \ref{tab:psyche_char} for a complete summary of Psyche]{Drummond2018}. Due to its spectral similarities to iron meteorites in the visible to near-infrared wavelength range of 0.4 – 2.5 $\mu m$, Psyche is traditionally thought to be predominantly metal \citep[][Figure \ref{fig:psyche_vnir}]{Chapman1973,Chapman1979}. Three main hypotheses interpret Psyche’s metallic composition. First, the long-standing view is that Psyche may be the stripped, remnant core of a differentiated, Vesta-size object \citep{Bell1989}. A more recent hypothesis suggests that Psyche’s metal-rich surface is a result of ferrovolcanism that protruded through a relatively thin mantle (25 km) as Psyche formed \citep{Johnson2020}. The third, alternative hypothesis describes Psyche as a collection of undifferentiated, highly reduced metal material that accreted near the Sun (first hypothesized by \citet{Bottke2006}, and attributed to Psyche by \citet{ElkinsTanton2016}). It is challenging to confirm one of these three hypotheses due to a range of possible meteorite types related to Psyche. The upcoming NASA Discovery-class Psyche orbiter will arrive at the asteroid to study its composition, topography, and the possible presence of a paleomagnetic field \citep{ElkinsTanton2014,ElkinsTanton2015}. These measurements, coupled with ground-based work, will shed insight into the mechanisms that formed Psyche and other similar bodies.

Various lines of evidence outside of reflectance spectroscopy have shown that the surface of Psyche is likely dominated by metal. The first radar observations of Psyche showed a radar albedo {(0.29 $\pm$ 0.11)} more than three times higher than normal S- and C-type asteroids \citep{Ostro1985}. Later observations of the object reaffirmed this high radar albedo, with \citet{Shepard2008,Shepard2017} attributing it to a surface dominated by iron and nickel. Using the Very Large Telescope (VLT) Mid-Infrared Interferometric Instrument (MIDI), \citet{Matter2013} found that Psyche has a high thermal inertia that is consistent with a surface primarily composed of metal. \citet{Kuzmanoski2002} observed the orbital perturbation of asteroid (13206) Baer on (16) Psyche, calculating a density of 6980 $\pm$ 580 kg m$^{-3}$ that is consistent with the density of iron meteorites studied on Earth (7500 kg m$^{-3}$). Later density estimates from radar yielded a density of 4500 $\pm$ 1400 kg m$^{-3}$, which can be attributed to an Fe-Ni composition with 40\% macroporosity \citep{Shepard2017}. More recent measurements have disputed this, with \citet{Siltala2020} obtaining a density estimate of $2360_{-980}^{+680}$  kg m$^{-3}$. While this data is within 3$\sigma$ of accepted literature values, it may indicate higher macroporosity and/or a lower metal content. The actual density of Psyche, as well as how its surface composition relates to its bulk composition, has implications for the Psyche mission. {This would include the ability of the \emph{Psyche} spacecraft to enter into the appropriate orbits around the asteroid, which depends upon its gravity field and hence, its density. Our study would also enable better interpretation of the remote sensing data from the \emph{Psyche} spacecraft.}

With the presence of a metal-rich surface established, more recent studies of Psyche have focused on identifying other minerals and their variation across the asteroid’s surface. A weak absorption feature was detected near 0.93 $\mu m$ \citep{Clark2004,Hardersen2005,OckertBell2010}, which \citet{Sanchez2017} attributed to a 6\% $\pm$ 1\% weight abundance of low-Fe pyroxene that may be mixed heterogeneously across the surface (Table \ref{tab:psyche_char}). Mid-infrared (1.9 – 4.2 $\mu m$) observations of Psyche also show a 3.0 $\mu m$ absorption feature that may indicate the presence of hydrated carbonaceous material \citep[][Table \ref{tab:psyche_char}]{Takir2017}. The presence of nonmetal material places constraints upon the formation and evolutionary history of Psyche. If Psyche was initially formed from differentiated material, it was likely impacted with exogenic carbonaceous materials. In such a case, carbonaceous material would lie as a thin veneer on the surface and may be distributed heterogeneously. 

The presence of three components (metal, pyroxene, and exogenic carbonaceous chondrite) in the surface regolith of Psyche presents us with an opportunity to revisit material abundance and macroporosity. We created three-component mixtures and obtained their visible to near-infrared (0.35 - 2.5 $\mu m$) spectra in the lab to identify the best matches to Psyche’s telescopic spectrum. Looking at the 0.93 $\mu m$ band depth and the VNIR spectral slope, we are able to put constraints on carbonaceous chondrite and low-Fe pyroxene content within our mixtures. We discuss the implications of our work on the composition and history of Psyche, which has broad implications for the formation of similar planetesimals in the main belt.

\section{Methods} \label{sec:methods}
\subsection{Sample Selection} 
\begin{figure}[ht!]
\plotone{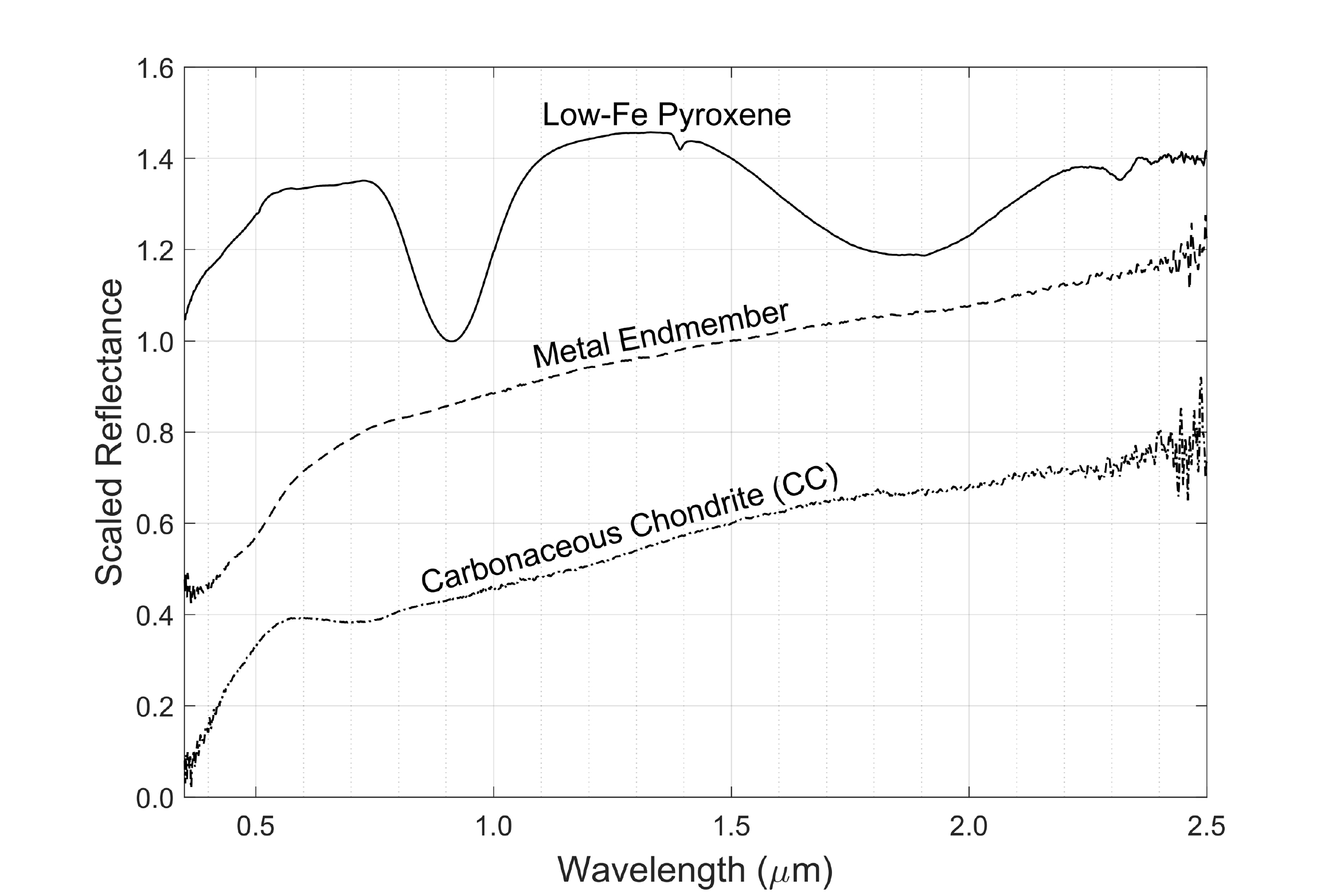}
\caption{Laboratory spectra of three endmembers (low-Fe pyroxene, metal, and carbonaceous chondrite) used in our study. Low-Fe pyroxene has two absorption bands at 0.93 and 2.0 $\mu m$ due to the presence of Fe$^{2+}$. The weak, sharp features at 1.4 and 1.9 $\mu m$ indicate minor amounts of hydrated clays that are common weathering products of pyroxene minerals. The metal component has a featureless spectrum while the carbonaceous chondrite has a slightly shallower slope and a weak absorption feature near 0.7 $\mu m$. The spectra are normalized to unity at 1.5 $\mu m$ and then offset by 0.4 to improve clarity.}
\label{fig:endmembers}
\end{figure}

Our three-component mixtures were a combination of relevant materials thought to be on the surface of Psyche: metal, low-Fe pyroxene, and carbonaceous chondrite (Figure \ref{fig:endmembers}). The metal endmember is a combination of two meteoritic powders from Gibeon and Georgetown. Gibeon, a common iron IVA meteorite, was discovered in Namibia in 1838 \citep[][see Table \ref{tab:endmembers} for endmember summary]{Moore1970}. Georgetown belongs to the IAB complex and was found in Australia in 1988 \citep{Weisberg2008}. Due to the difficulty of grinding down metal meteorites, Gibeon powder was obtained as cutting shavings contaminated with cutting oil. The powder was repeatedly washed with acetone until there was no residual odor in the sample caused by the oil. The purity of the sample was confirmed by the absence of hydrocarbon absorptions measured in its VNIR spectrum. It was subsequently sieved with acetone to select grain sizes between 106 and 212 $\mu m$ and dried in a heated vacuum oven. We selected Gibeon as a metal endmember due to its featureless, red-sloped spectrum and its relative ease to acquire. Since Gibeon’s VNIR slope is too steep to directly match Psyche’s spectrum, Georgetown powder was added to decrease the slope and better match Psyche. Georgetown powder was acquired by hand-crushing a small fragment and extracting less reflective metallic portions of this slab, which are more fragile. These fragments were crushed using a mortar and pestle and dry sieved to select for grain sizes less than 45 $\mu m$. It is unclear as to why these fragments of Georgetown were different from the surrounding texture, but spectra of this powder revealed a featureless slope shallower than Gibeon’s slope, making it ideal for mixing. The shallower spectral slope of Georgetown may be due to the inclusion of more sulfides in the form of troilite. Our low-Fe pyroxene comes from a terrestrial sample of enstatite from Mirabel Springs, California, that was crushed with a mortar and pestle and dry sieved to less than 45 $\mu m$. Weak absorptions in the VNIR spectrum at 1.4 ($\sim$ 3\% depth) and 1.9 ($\sim$ 0.2\% depth) $\mu m$ indicate minor amounts of hydrated clays, a common weathering product of pyroxene minerals. Our carbonaceous chondrite originates from Murchison, a prominent CM2 meteorite that was observed to fall in Australia in 1969 (C.B.M. 1970). This endmember was obtained as leftover material from previous laboratory experiments and is representative of CM2 material. Like our low-Fe pyroxene, it was crushed with a mortar and pestle and constrained to less than 45 $\mu m$ with a sieve.

The analysis of laboratory mixtures can never provide a perfect match to a complex, remote object. However, the careful selection of our three endmembers enables us to set a benchmark for future observations. The combination of Gibeon and Georgetown meteoritic powder simply allowed a featureless spectrum that matched the slope of Psyche. The larger grain size of Gibeon (106 – 212 $\mu m$) from our other endmember materials ($<$45 $\mu m$) will likely cause subtle spectral variations, primarily in spectral slope. Since Gibeon does not contribute to the 1 and 2 $\mu m$ absorption bands, we do not expect this grain size difference to have a major impact on our analysis. While our Mirabel Springs pyroxene matches the low-Fe pyroxene observed on Psyche, we were mainly focused on incorporating an endmember that introduced the 0.93 $\mu m$ absorption feature. Other pyroxene endmembers, including those with higher Fe content, have similar band depths and would not significantly affect our results. Lastly, our selection of CM2 as the carbonaceous chondrite endmember was due to its presence on the surface of other large main belt asteroids, such as Vesta, and in HED meteorites \citep[e.g.,][]{Reddy2012}. Clasts of CM2 chondrites have been found in a wide variety of solar system materials, attesting to its high availability and ability to be transported. Other than a weak absorption feature near 0.7 $\mu m$, carbonaceous chondrite also has a relatively featureless slope and largely acts as a darkening agent. From this, we consider our results to be fairly robust in predicting the material content, regardless of the exact selection of endmembers.

\subsection{Sample Preparation} 
\begin{figure}[ht!]
\plotone{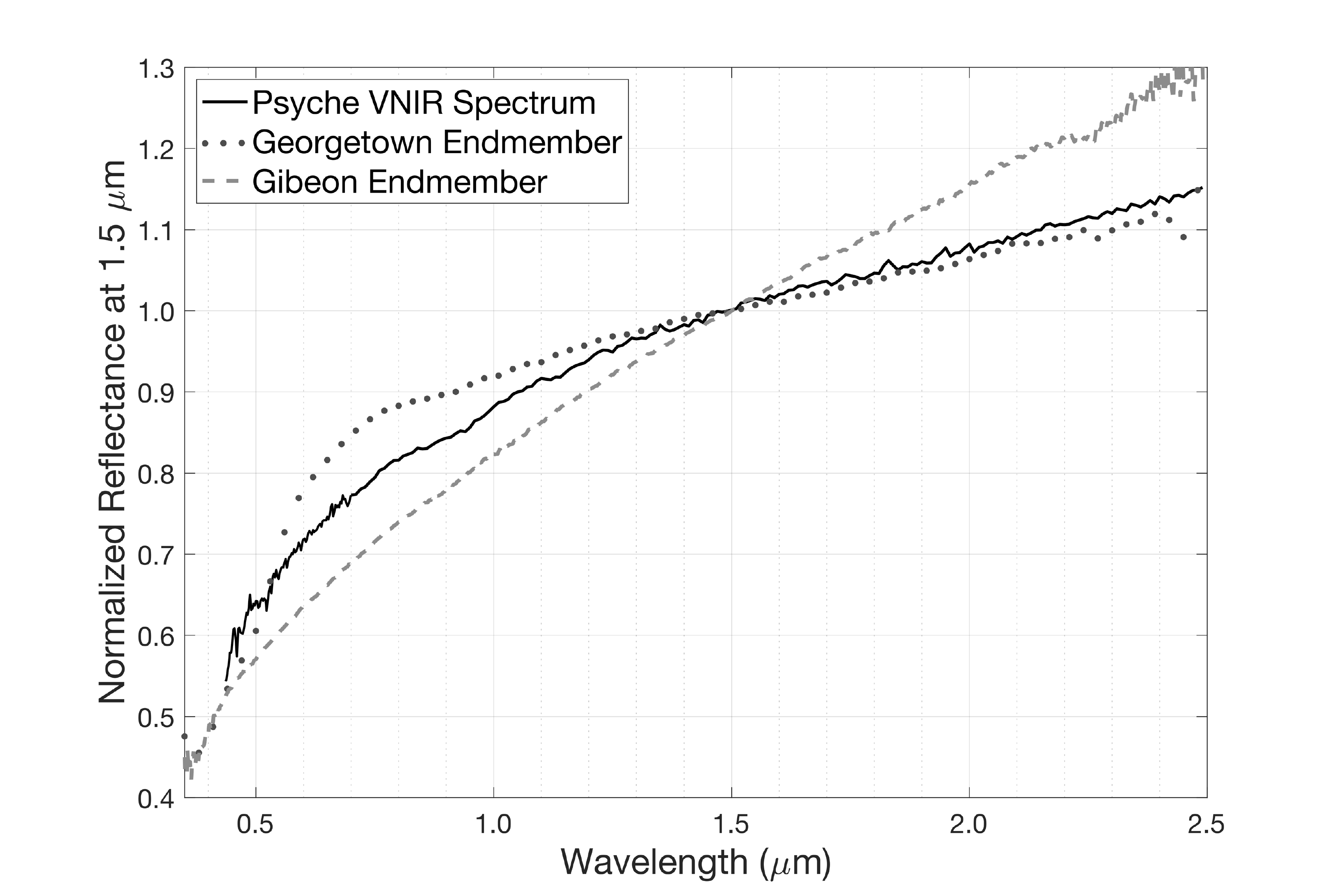}
\caption{Laboratory spectra of the two metal meteoritic powders, Georgetown and Gibeon. While Georgetown’s spectral slope is too shallow to match the telescopic spectrum of asteroid (16) Psyche, Gibeon’s spectral slope is too steep. The two endmembers were mixed together to create a better spectral analog to Psyche (Figure  \ref{fig:metal_mix}). The data are normalized to unity at 1.5 $\mu m$.}
\label{fig:metal_endmembers}
\end{figure}

\begin{figure}[ht!]
\plotone{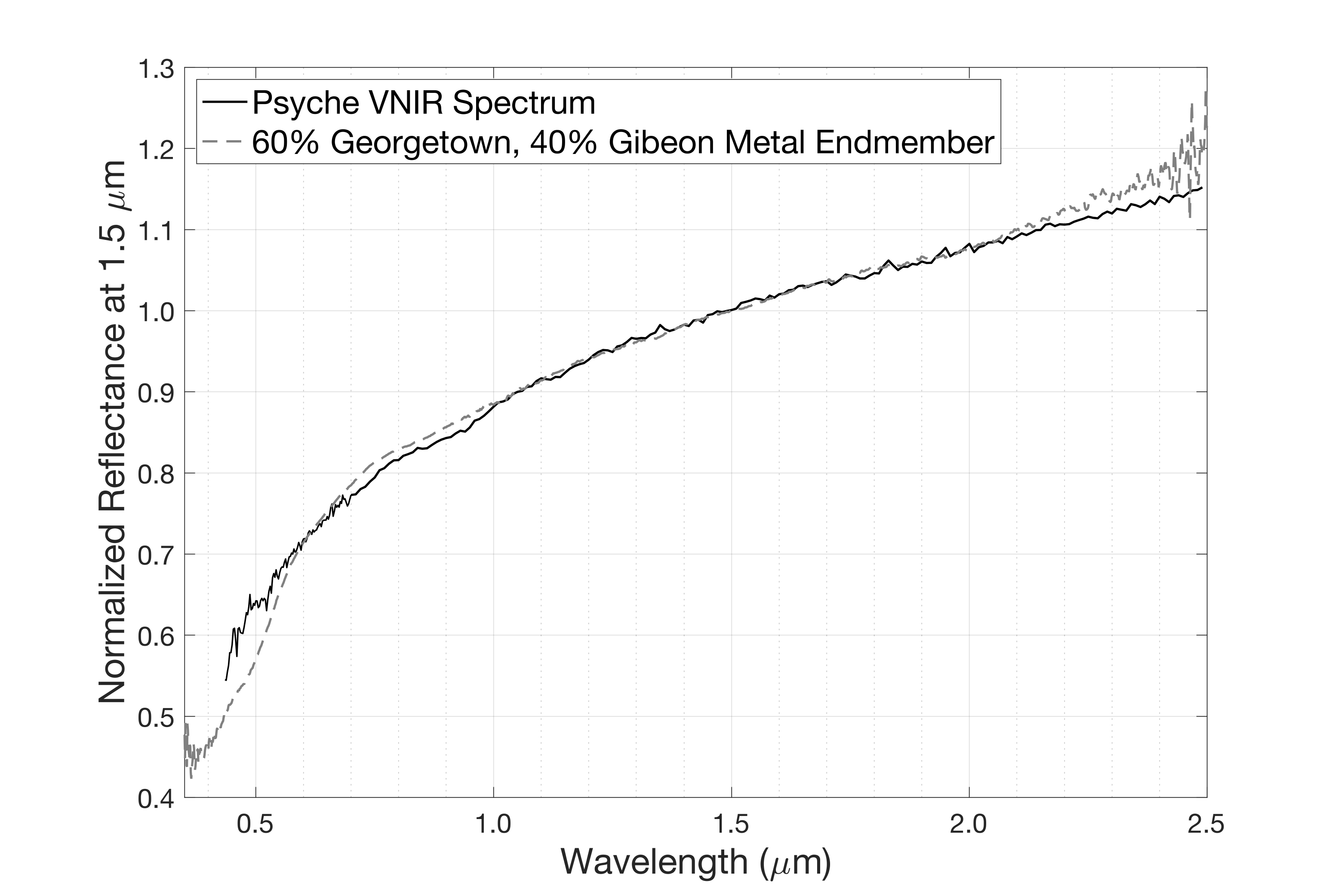}
\caption{Laboratory spectrum of our two-component mixture of 60\% Georgetown and 40\% Gibeon plotted against the VNIR spectrum of Psyche. The 60 to 40 ratio of Georgetown to Gibeon provided the best spectral slope match to Psyche's spectrum. The data are normalized to unity at 1.5 $\mu m$.}
\label{fig:metal_mix}
\end{figure}

Due to the low abundance of accessible Georgetown material, certain measures were taken to conserve it throughout our mixtures. While creating the metal endmember from Gibeon and Georgetown (Figure \ref{fig:metal_endmembers}), we started with a fixed amount of Georgetown material and added in progressive amounts of Gibeon. After mixing and measuring the mixture’s spectral slope, further additions of Gibeon were made to achieve a new mixture that could accurately match Psyche’s broadband slope (Figure \ref{fig:metal_mix}). We repeated this process to maximize the number of mixtures created while conserving the limited meteoritic material. We found that the spectrum of 60\% Georgetown and 40\% Gibeon is the best spectral slope match to Psyche’s spectrum.

While this mixing process is straightforward to apply to two-component mixtures, three-component (tertiary) mixtures with metal, pyroxene, and carbonaceous chondrite are more complex to systematically vary. For our tertiary mixtures, we started with a fixed amount of metal material. In our first three series, we then added in a fixed amount of low-Fe pyroxene (3, 5, or 7 wt.\%). To gauge the effects of adding in carbonaceous chondrite, small amounts of Murchison were then added incrementally to the mixtures, effectively reducing metal content. This was done in a process we referred to as “repeated additive mixing” (RAM). While RAM could be used to conserve the original metal and pyroxene material as an experiment progressed, there were more deviations in the integer weight percent of the low-Fe pyroxene as more carbonaceous chondrite was added. For example, the addition of carbonaceous chondrite in the early iterations of Series 3 resulted in our low-Fe pyroxene wt.\% to drop from 7\% to 6.5\% as a result of the total mixture mass increasing. To correct this, small amounts of low-Fe pyroxene were added approximately every five experimental runs to maintain our targeted weight ratios. In Series 4, low-Fe pyroxene was incrementally added to a fixed amount of metal and carbonaceous chondrite to explore its effects. RAM was again used to conserve this material, with carbonaceous chondrite occasionally being added to retain our intended mixture ratios. 

For the RAM technique to be valid, we must assume that the total weight and composition of the mixture stays constant during the preparation process. To test this, the mixtures were weighed before and then multiple times throughout each experimental series. On average, a mixture lost 0.5\% of its total weight as a result of transferring it between containers throughout each experimental run. A single run consists of the sample being weighed, prepared in a sample cup, transferred to the spectrometer, and returned to its original container. During our experimental series, total weights were consistently taken of the mixture to revise our RAM calculations. Assuming that the mixture’s mass was lost homogeneously, these revisions to our calculations were largely insignificant.

Mixture and component weights were all measured with a calibrated precision balance with a precision of $\pm$ 0.0001 g. Typical additions of carbonaceous chondrite and pyroxene were in the range of 0.0050 g (2\% uncertainty). The small, infrequent corrections of the corresponding, alternate material (e.g., adding a small amount of pyroxene in Series 1-3, adding a small amount of carbonaceous chondrite in Series 4) were close to 0.0010 g (10\% uncertainty). Variations in these additions were due to changes in the total mixture weight and the ratio of components. 

\subsection{Data Collection and Analysis}
All laboratory VNIR spectra were acquired with an ASD LabSpec 4 Hi-Res spectrometer. This spectrometer enables 3 nm resolution at 0.70 $\mu m$ and 6 nm resolution at 1.40 and 2.10 $\mu m$. A 120 Watt quartz-tungsten bulb was used as the light source at an incidence angle of 0$^\circ$, while emission was measured at an angle of 30$^\circ$. Spectra were measured relative to a baseline Spectralon disk and corrected for dark current. Data were then processed with a python script that removed a known infrared feature of Spectralon and corrected for any detector offset at 1.0 and 1.8 $\mu m$ \citep{Kokaly2017}. All spectra were then normalized to unity at 1.5 $\mu m$. 

Once data were reduced, spectral slope and absorption band properties could be measured. The 0.93 $\mu m$ band depths were calculated with a python script described in \citet{Sanchez2017}. First, two continuum points ($\sim$0.75 and $\sim$1.1 $\mu m$) were selected. Then, after dividing out the linear continuum, two points surrounding the band center were chosen to constrain a fourth-order polynomial fit. Since the spectral reflectance errors for our laboratory mixtures are very small, band parameter uncertainties are dominated by the selection of the spectral continuum. To account for this source of uncertainty, this band analysis procedure was repeated five times per spectrum, using slightly different selections of continuum points, to characterize the variability introduced by removing the background slope. The mean values of band depth are given in Table \ref{tab:mixes}, with uncertainties taken as the standard deviation of the five measurements. The spectral slope of each sample was calculated using 1.8 and 0.8 $\mu m$ reflectance values. These areas of the spectrum were relatively unaffected by band depths and were also used by \citet{Cloutis2010} to analyze the spectral slopes of metallic surfaces.

\section{Results} \label{sec:results}

Our laboratory experiments consisted of four series that explored the effects of adding low-Fe pyroxene and carbonaceous chondrite to meteoritic metal. Details of the mixtures are presented in Table \ref{tab:mixes}. Series 1-3 observed the effects of incrementally adding carbonaceous chondrite to mixtures of fixed low-Fe pyroxene content (3, 5, and 7 wt.\%) and meteoritic metal. Series 4 explored the effects of incrementally adding low-Fe pyroxene to a fixed amount of carbonaceous chondrite (predicted by Series 3) and meteoritic metal. Band-depth data and overall reflectance were used to interpret these additions of material and constrain their abundance.

\subsection{Series 1: Fixed Low-Fe Pyroxene (3 wt.\%)}
Our first experimental series tested mixtures containing 3\% low-Fe pyroxene, a value that is less than the minimum low-Fe pyroxene content predicted to be on Psyche’s surface by \citet{Sanchez2017}. Before any carbonaceous chondrite was added, PMix 2 (97\% metal, 3\% low-Fe pyroxene) had a 0.93 $\mu m$ band depth of 0.34\% $\pm$ 0.21\%. Adding in a small amount (3\%) of carbonaceous chondrite resulted in a band depth of 0.26\% $\pm$ 0.15\%, slightly lower than PMix 2, but still within error margins. At this point, no further mixtures with 3\% low-Fe pyroxene were made. Through this quick series, it was clear that more than 3\% low-Fe pyroxene was needed in our mixtures to best match the 0.93 $\mu m$ band depth (see Table \ref{tab:mixes} for a complete summary). With this, the effect of carbonaceous chondrite remained unconstrained.

\subsection{Series 2: Fixed Low-Fe Pyroxene (5 wt.\%)}
Series 2 explored tertiary mixtures with 5\% low-Fe pyroxene, a value within the limits of what \citet{Sanchez2017} predicted. Before adding carbonaceous chondrite, FMix 1 (95\% metal, 5\% low-Fe pyroxene) had a moderate band depth of 1.24\% $\pm$ 0.15\%, which is a plausible analog to the surface of Psyche. Following the addition of a small amount of carbonaceous chondrite (1\%), the band depth of FMix 2 (94\% metal, 5\% low-Fe pyroxene, 1\% carbonaceous chondrite) quickly dropped to 0.50\% $\pm$ 0.12\%. This band depth was weakened slightly further in FMix 3 (92\% metal, 5\% low-Fe pyroxene, 3\% carbonaceous chondrite) to 0.46\% $\pm$ 0.24\% (results are summarized in Table \ref{tab:mixes}).

\subsection{Series 3: Fixed Low-Fe Pyroxene (7 wt.\%)}
\begin{figure}[ht!]
\plotone{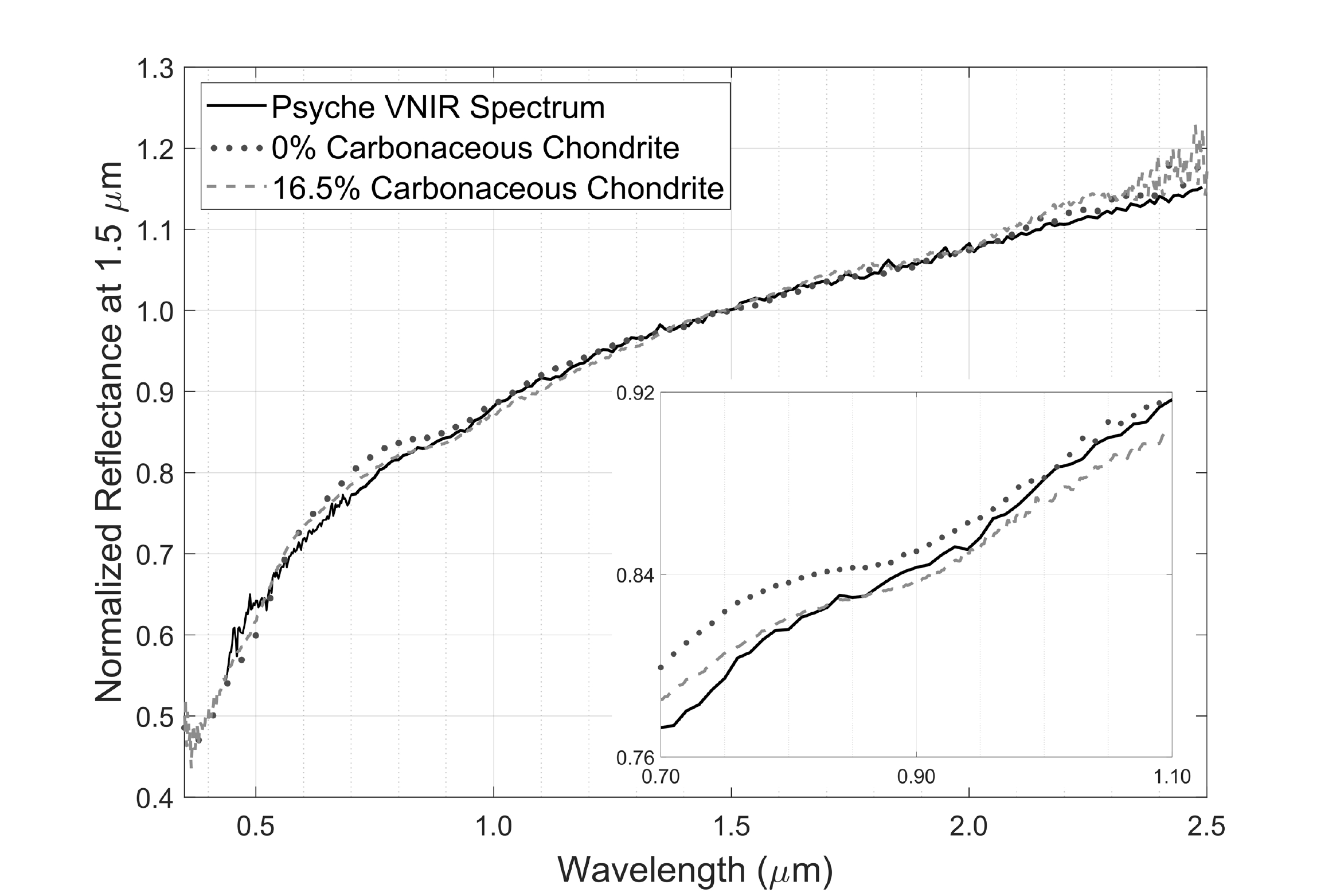}
\caption{Laboratory spectra of mixtures with constant 7\% low-Fe pyroxene content and increasing carbonaceous chondrite (selected from Series 3). The $\sim$ 0.7-1.2 $\mu m$ region of the spectra decreases in normalized reflectance as more carbonaceous chondrite is added due to reduction in overall reflectance. An inset plot from 0.7 to 1.1 $\mu m$ is included for clarity. The data are normalized to unity at 1.5 $\mu m$.}
\label{fig:cc_spectra}
\end{figure}

Series 3 included tertiary mixtures with 7\% low-Fe pyroxene, a value in the upper limits of what \citet{Sanchez2017} predicted. With no carbonaceous chondrite, BMix 2 (93\% metal, 7\% low-Fe pyroxene) had a measured 0.93 $\mu m$ band depth of 1.74\% $\pm$ 0.25\%. This is significantly higher than the measured 1.3\% $\pm$ 0.1\% band depth by \citet{Sanchez2017} corresponding to 6\% pyroxene. This would imply that 7\% low-Fe pyroxene content is too high, or there is another component that weakens this band. Following this, multiple amounts of carbonaceous chondrite were added using RAM with little effect on band depth. It was not until BMix 8 (85.5\% metal, 7\% low-Fe pyroxene, 7.5\% carbonaceous chondrite) that any noticeable change in band depth was observed. Even in this mixture, band depth was only reduced to 1.56\% $\pm$ 0.21\%, a value still too high to match Psyche. It is unclear why these first additions of carbonaceous chondrite in this series were not as effective in reducing 0.93 $\mu m$ band depth in Series 2. The increased amount of low-Fe pyroxene (5\% in Series 2 versus 7\% in Series 3) may result in a more carbonaceous chondrite needed to effectively “mask” the 0.93 $\mu m$ absorption feature. Band depth continued to decrease with added carbonaceous chondrite, creating BMix 12 (82.5\% metal, 7\% low-Fe pyroxene, 10.5\% carbonaceous chondrite) that matches the measured band depth of Psyche at 1.32\% $\pm$ 0.22\%. More carbonaceous chondrite was added incrementally to gauge an upper limit. BMix 15 (76.5\% metal, 7\% low-Fe pyroxene, 16.5\% carbonaceous chondrite) had a measured 0.93 $\mu m$ band depth of 1.18\% $\pm$ 0.11\%, a value that is slightly lower than the 1.3\% $\pm$ 0.1\% limits that \citet{Sanchez2017} predicted (see Table \ref{tab:mixes} for full band-depth data). Assuming 7\% low-Fe pyroxene content, enough ($\sim$8 to $\sim$17 wt.\%) carbonaceous chondrite must be present to mask the 0.93 $\mu m$ band depth of Psyche. As shown in Figure \ref{fig:cc_spectra}, adding carbonaceous chondrite to the mixtures also results in an overall decrease in reflectance in the 0.7 – 1.2 $\mu m$ region of the spectrum. Along with matching the $\sim$1.3\% band depth, BMix 12 had the correctly scaled albedo near the center of this region at 0.9 $\mu m$. To verify that this mixture could match Psyche, it is important to compare its absolute reflectance with the albedo of Psyche. For each mixture, we analyzed the 0.55 $\mu m$ absolute reflectance, which is a proxy for albedo (Table \ref{tab:mixes}). The absolute reflectance of BMix 12 at 0.55 $\mu m$ is 0.069, which is within the uncertainty of the geometric albedo of Psyche (0.093 $\pm$ 0.024) measured by \citet{Mainzer2016}.

\subsection{Series 4: Fixed Carbonaceous Chondrite (10.5 wt.\%)}
\begin{figure}[ht!]
\plotone{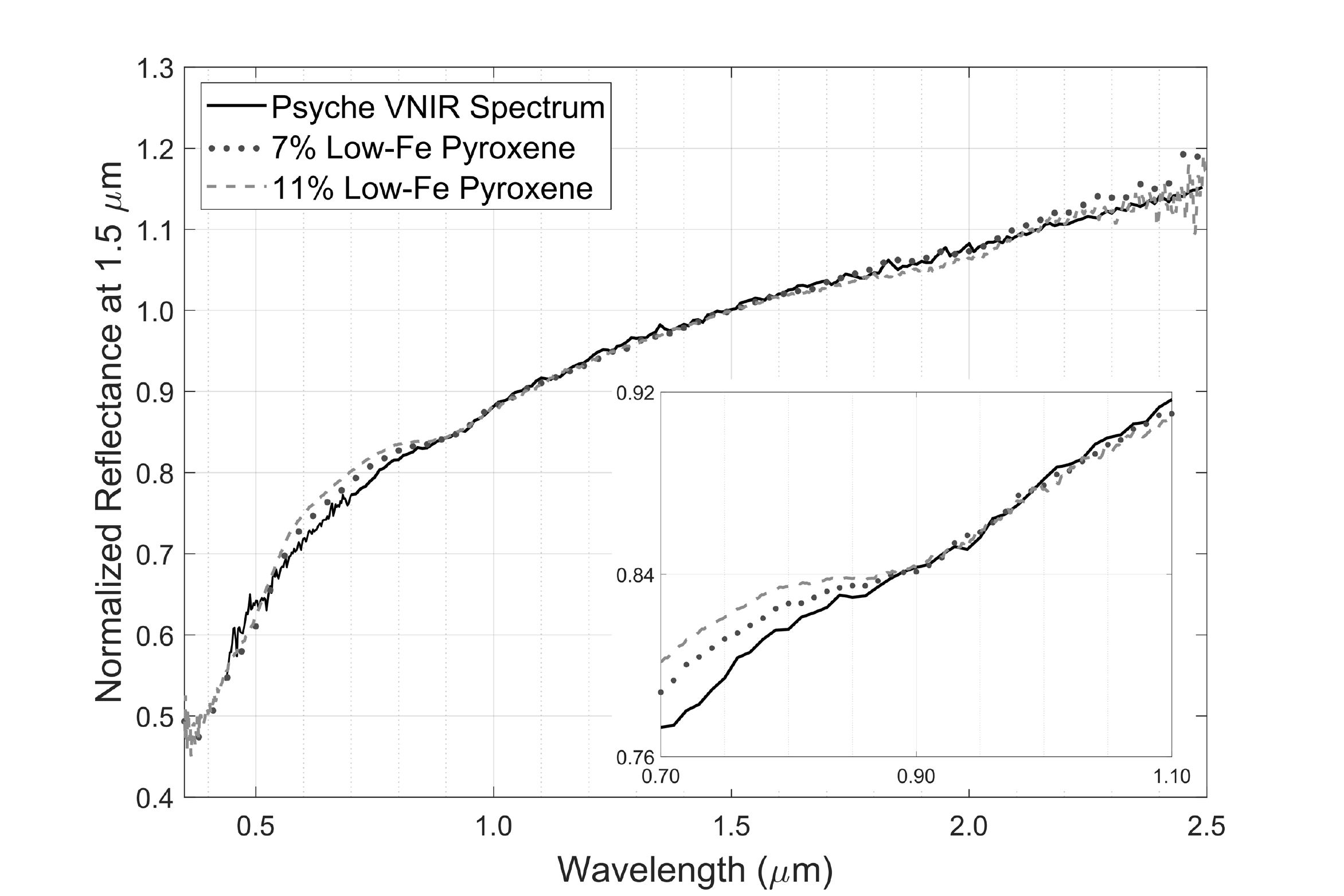}
\caption{Laboratory spectra of mixtures with constant 10.5\% carbonaceous chondrite and increasing low-Fe pyroxene content (selected from Series 4). The $\sim$ 0.6–0.9 $\mu m$ region of the spectra gradually increases in reflectance as more pyroxene is added. An inset plot from 0.7 to 1.1 $\mu m$ is included for clarity. The data are normalized to unity at 1.5 $\mu m$.}
\label{fig:px_spectra}
\end{figure}

After exploring the effects of carbonaceous chondrite in our mixtures and setting its upper limit, we sought to do the same with low-Fe pyroxene content. It is understood that adding low-Fe pyroxene results in a deeper 0.93 $\mu m$ band depth, as seen from the two-component mixtures in our first three series (PMix 2, FMix 0, and BMix 2). With a fixed, plausible amount of carbonaceous chondrite from BMix 12 ($\sim$10.5\%), Series 4 explored the quantitative and graphical effects of adding low-Fe pyroxene to set an upper limit. Starting with BMix 12, low-Fe pyroxene was added in increments of 1\% using RAM, effectively decreasing metal content. As shown in Table \ref{tab:mixes}, increasing low-Fe pyroxene results in a deeper 0.93 $\mu m$ band depth. FMix 7, 8, and 9 all increase in band depth but may still be within the upper limits of what \citet{Sanchez2017} predicted. It is not until FMix 10 (78.5\% metal, 11\% low-Fe pyroxene, 10.5\% carbonaceous chondrite) that the measured band depth reaches 1.86\% $\pm$ 0.23\% (Table \ref{tab:mixes}), a value much too high to match Psyche. Graphically, the addition of low-Fe pyroxene with mixtures already containing metal and carbonaceous chondrite results in an increase in reflectance from $\sim$ 0.6 to 0.9 $\mu m$ (Figure \ref{fig:px_spectra}). With this, low-Fe pyroxene content above 10\% begins to introduce a band II (1.9 $\mu m$) absorption feature that is not present in Psyche’s spectrum.

\subsection{Summary of Results}
Trends in band depths and normalized reflectance were used to help constrain the upper and lower limits of low-Fe pyroxene and carbonaceous chondrite. From analyzing 0.93 $\mu m$ band depths, Series 1 and 2 illustrate the need for more than 5\% low-Fe pyroxene assuming the presence of carbonaceous chondrite. Series 4 also utilizes band depths to show that low-Fe pyroxene content above 10\% results in 0.93 $\mu m$ band depths that are far too great, as well as the introduction of the band II absorption. Graphically, Series 4 shows a preference for a lower amount of low-Fe content to better match Psyche’s spectra. Our results show that 7 wt.\% low-Fe results in the best fit for both band depth (Table \ref{tab:mixes}) and 0.6 – 0.9 reflectance (Figure  \ref{fig:px_spectra}). With this, we constrain low-Fe pyroxene to within $\sim$ 5 – 10 wt.\%, assuming the presence of carbonaceous chondrite. 

\begin{figure}[ht!]
\plotone{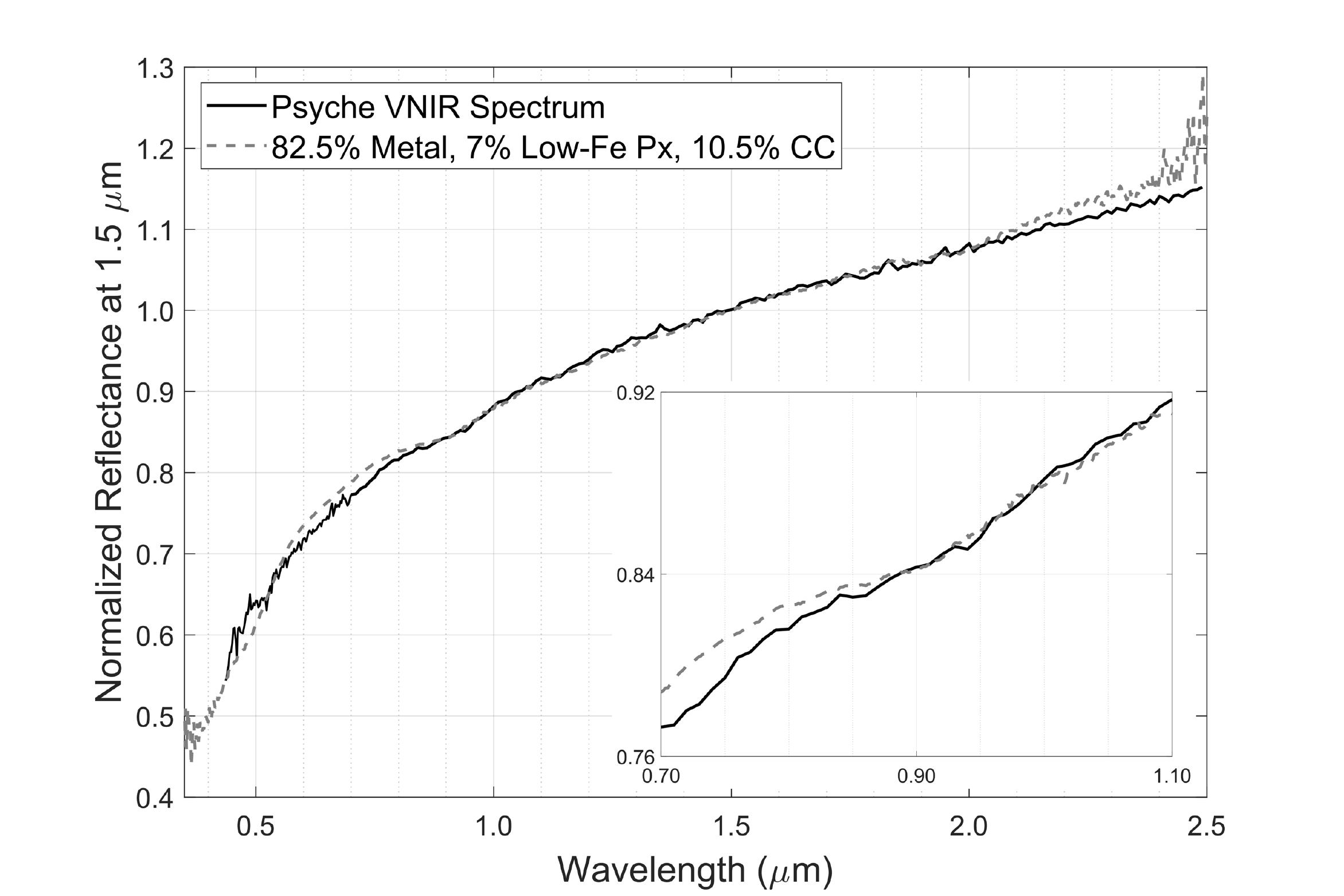}
\caption{Laboratory spectrum of mixture BMix 12 (82.5\% metal, 7\% low-Fe pyroxene, 10.5\% carbonaceous chondrite) that best matches the telescopic spectrum of asteroid Psyche. This laboratory spectrum has the same spectral slope and band depth as the asteroid. An inset plot from 0.7 to 1.1 $\mu m$ is included for clarity. The data are normalized to unity at 1.5 $\mu m$.}
\label{fig:bmix12}
\end{figure}

Through our first three series, it is clear that carbonaceous chondrite can effectively reduce the 0.93 $\mu m$ band depth induced by low-Fe pyroxene. The amount of carbonaceous chondrite needed to mask the low-Fe pyroxene 0.93 $\mu m$ feature is dependent on the amount of low-Fe pyroxene present. Graphically, Series 3 shows the addition of carbonaceous chondrite results in a reduction in normalized reflectance between 0.7 and 1.2 $\mu m$ (Figure  \ref{fig:cc_spectra}). This has many implications in terms of constraining composition. Considering only band depths, one may wonder if increasing carbonaceous chondrite can continue to mask increasing levels of low-Fe pyroxene. Including reflectance, we can conclude that this is not the case. There is an upper limit of $\sim$17\% carbonaceous chondrite where the mixture’s normalized reflectance near 1 $\mu m$ dips too low to accurately match Psyche. Adding more low-Fe pyroxene in an attempt to re-deepen the 0.93 $\mu m$ band depth will only increase the normalized reflectance from 0.6 to 0.9, straying further from Psyche’s spectra and still not raising the 1 $\mu m$ region’s albedo. If material analogous to carbonaceous chondrites is present on the surface of Psyche, we predict its content to be $\sim$ 10.5\%, which accurately matches the 1.3\% band depth (Table \ref{tab:mixes}) and the relative albedo at 0.9 $\mu m$ (Figure \ref{fig:bmix12}) with 7\% low-Fe pyroxene. We constrain carbonaceous chondrite content to within $\sim$ 8 – 17 wt.\%, the range of values within limits of the predicted 0.93 $\mu m$ band depth of 1.3\%. A summary plot showing our three-component laboratory mixtures of metal (Gibeon/Georgetown), low-Fe pyroxene, and carbonaceous chondrite (Murchison) is shown in Figure \ref{fig:ternary_plot} to help visualize the mixture ratio space that we explored as part of our study.

\section{Discussion} \label{sec:discussion}

Our three-component laboratory mixtures are an important step in constraining the detailed surface composition of asteroid (16) Psyche and offering insight into its formation prior to the arrival of the Psyche spacecraft. The long-standing view of Psyche’s formation describes a differentiated object that was stripped of its mantle and crust, leaving a remnant metal core seen with radar \citep{Ostro1985,Shepard2008,Shepard2017} and spectroscopic observations \citep{Bell1989,Gaffey1993}. Numerical simulations have demonstrated that these hit-and-run collisions can expose the metal core of differentiated objects \citep{Asphaug2006,Asphaug2014}, though it is unlikely that these events would strip all mantle and crustal material \citep{Hardersen2005,Sanchez2017}. Most planetesimals undergoing differentiation produce an olivine mantle as the first result of crystallization, making it the assumed remnant silicate material on Psyche \citep{ElkinsTanton2013}. The apparent lack of olivine’s weak absorption feature centered at 1.05 $\mu m$ is one of the more puzzling aspects of Psyche’s spectra. Instead, observations indicate a weak ($\sim$1.3\%) absorption feature at 0.93 $\mu m$ \citep{Sanchez2017} that we attribute to $\sim$7 wt.\% low-Fe pyroxene. 

The presence of pyroxene instead of olivine must be accounted for by Psyche’s formation history, which was explored in four main mechanisms described by \citet{Hardersen2005}. First, the pyroxene on Psyche’s surface may represent residual mantle material following the collisions that transformed its parent body into a remnant core. This scenario is consistent with many early models for Psyche’s formation, but does not directly explain the absence of olivine on the surface. A second mechanism suggested by \citet{Hardersen2005} describes a smelting option that incorporates carbon as a reducing agent. In this high-temperature reaction, fayalite (olivine) and carbon can react to form enstatite, metallic iron, and possibly silica \citep{Walker1993}. While this reaction may have been possible due to the presence of carbonaceous chondrite material, the small amount of iron in enstatite that is observed on the surface of Psyche means the reaction did not run to completion. Like the previous scenario, this mechanism is difficult to explain because it assumes there is residual olivine material on the surface. \citet{Takir2017} also note that the carbonaceous chondrite material is hydrated, suggesting its emplacement after Psyche differentiation. A third option explored by \citet{Hardersen2005} places the primitive, reduced CB chondrites as analogs to Psyche. These meteorites contain $\sim$40\% low-Fe silicate and $\sim$60\% metal clasts \citep{Krot2005}, though their delivery to Psyche and other M-types is assumed to be from collisions. It is unlikely that these collisions, which are mostly thought to be localized, can result in a surface that is largely homogeneous in silicate content.

A recently proposed formation hypothesis does not require the stripping of Psyche’s mantle and crust. Instead, Psyche may have been a differentiated object that began to cool from the outside in. The molten iron and nickel, propagating upwards through the mantle in a series of dykes, may have erupted over the surface, creating a mixture of metal and silicate crustal material that could be further modified by collisions \citep{Johnson2020}. During Psyche’s early geologic history, it is possible that high core temperatures (1261–1811 K) may have resulted in an excess pressure of millions of pascals, allowing dyke propagation up to 50 km in height \citep{Johnson2020}. Another formation hypothesis describes an unusual initial environment without differentiation. If a planetesimal had a low-pressure interior and a bulk composition dominated by reduced, silicate-rich material, then pyroxene would be the first mineral to crystallize \citep{ElkinsTanton2013}. This hypothesis could explain the lack of olivine and the unique presence of pyroxene on Psyche’s surface. In this formation scenario, Psyche may have formed as an undifferentiated body near the Sun that may not have needed large impactors to strip its crust \citep{Weidenschilling1978,Bottke2006,ElkinsTanton2016}.

While it can be challenging to scale small meteorite samples to asteroids hundreds of kilometers in size, meteorites can still be relevant in setting constraints in composition and density. One formation scenario places Psyche as the possible parent body of mesosiderites, a type of stony-iron meteorite \citep{Mason1972}. Unlike pallasite meteorites, which are rich in olivine and thought to be from the core-mantle boundary of a differentiated object, mesosiderites contain mostly core (Fe-Ni metal) and crustal silicate (pyroxene) material \citep[][p. 19]{Weisberg2006}. \citet{Fieber-Beyer2011} identified the Maria asteroid family near the 3:1 Kirkwood Gap as a plausible source of mesosiderites; though mesosiderites share many spectral similarities to Psyche as well, including the 0.93 $\mu m$ pyroxene absorption feature \citep{Cloutis2010}. Using a numerical model, \citet{Scott2001} detailed a possible formation hypothesis that may explain the reaccretion of silicate material and lack of olivine on Psyche’s surface following a large impact. While this could provide the missing link between mesosiderites and their parent body, their compositions do not match directly. Mesosiderites are observed to have high-Fe silicate material (like that of asteroid Vesta); though Psyche has much lower iron content in its pyroxene \citep{Sanchez2017,ElkinsTanton2020}. Enstatite chondrites have also been considered as a possible analog to Psyche and other M-type asteroids. The lack of iron in enstatite chondrite’s pyroxene component, however, results in a featureless spectrum without the 0.9 $\mu m$ absorption feature. This effectively rules out enstatite chondrites as analogous material to Psyche and other M-types with low-Fe pyroxene content \citep{Hardersen2005}. While Psyche \citep[3,780 $\pm$ 340 kg m$^{-3}$,][]{ElkinsTanton2020} may be similar in density and composition to mesosiderites (3,100 – 7,200 kg m$^{-3}$) and enstatite chondrites \citep[3,100 – 3,800 kg m$^{-3}$,][]{Britt2003}, there is still no definitive meteorite analog to the asteroid.

Psyche is often thought of as being predominantly metal, but recent density estimates paint a more complicated picture. While a completely metallic composition would have a density upwards of 8,000 kg m$^{-3}$ , modern estimates of Psyche’s density are less than half of that, in the range of 3,780 $\pm$ 340 kg m$^{-3}$ 
\citep{ElkinsTanton2020}. To compare our results with other authors and simplify calculations regarding density and bulk porosity of Psyche, we have converted the material content of our best spectral match to volume percent. By volume, BMix 12 is 61.1\% metal, 12.5\% low-Fe pyroxene, and 26.4\% carbonaceous chondrite. This is assuming a density of metal (kamacite) as 7,870 kg m$^{-3}$ \citep{ElkinsTanton2020}, low-Fe pyroxene (enstatite) as 3270 kg m$^{-3}$ \citep{ElkinsTanton2020}, and carbonaceous chondrite (Murchison, CM2) as 2,310 kg m$^{-3}$ \citep{Macke2011}. Assuming no porosity, this laboratory mixture has a density of 5,827 kg m$^{-3}$, far too high to match recent density estimates of Psyche. To match the density of 3,780 kg m$^{-3}$ suggested by \citet{ElkinsTanton2020}, a total mixture composition of 39.6\% metal, 8.1\% low-Fe pyroxene, 17.1\% carbonaceous chondrite (25.2\% non-metal), and 35.1\% void space is needed. 

This high predicted porosity may be unusual for such a large object; though it is a similar value to what \citet{Shepard2017} predicted would be necessary ($\sim$40\%). While small bodies ($<$1 km in diameter) like Ryugu can have bulk porosities greater than 50\% \citep{Watanabe2019}, one would expect larger objects to have less porosity due to increased surface gravity. Using data from the upper crust of the Moon, which has an average porosity of 12\% \citep{Wieczorek2013}, \citet{ElkinsTanton2020} predicted a maximum bulk porosity of 20\% for small bodies. Applied to our best-fitting mixture, this porosity would yield a density of 4,662 kg m$^{-3}$, a value still too high to match recent estimates. Assuming that Psyche is composed of a similar set of three components, our predicted porosity (35\%) may indicate that larger asteroids can still retain relatively high bulk porosity despite increased surface gravity. One possible explanation of Psyche’s high porosity is that it may be a rubble pile similar to much smaller asteroids like Ryugu and Bennu. In such a case, Psyche would be much more heavily modified than previously thought, requiring a complete rework of formation scenarios. More work is needed to constrain the density of Psyche and other small bodies.

Studying the carbonaceous material content of asteroids can offer insight into their formation, such as detailing initial conditions and collisional history. Our constraints of carbonaceous chondrite on the surface of Psyche can help characterize M-type asteroids and how they differ from other bodies. Our results are in agreement with \citet{Rivkin2000}, who predicted that M-types larger than 65 km are likely to be hydrated. Psyche, the largest M-type asteroid with an average diameter of 223 $\pm$ 4 km \citep{Drummond2018}, certainly matches this criterion with our predicted carbonaceous chondrite abundance of 10.5 wt.\%. This agrees with the results of \citet{Takir2017}, who first observed the possible presence of hydrated material due to a 3.0 $\mu m$ absorption feature in Psyche’s NIR spectrum. While our work only focused on understanding the VNIR (0.35 – 2.5 $\mu m$) spectrum, we successfully studied the subtle effects carbonaceous chondrite had in reducing the 0.93-$\mu m$ band depth and affecting the 0.7 – 1.2 $\mu m$ region of our mixtures. Our results placed carbonaceous chondrite at a higher value than initially suspected. Our predicted value of $\sim$10.5 wt.\% indicates that Psyche’s regolith has more hydrated material and is less metallic than originally proposed.

\section{Summary} \label{sec:summary}

\begin{figure}[ht!]

\epsscale{0.85} 
\plotone{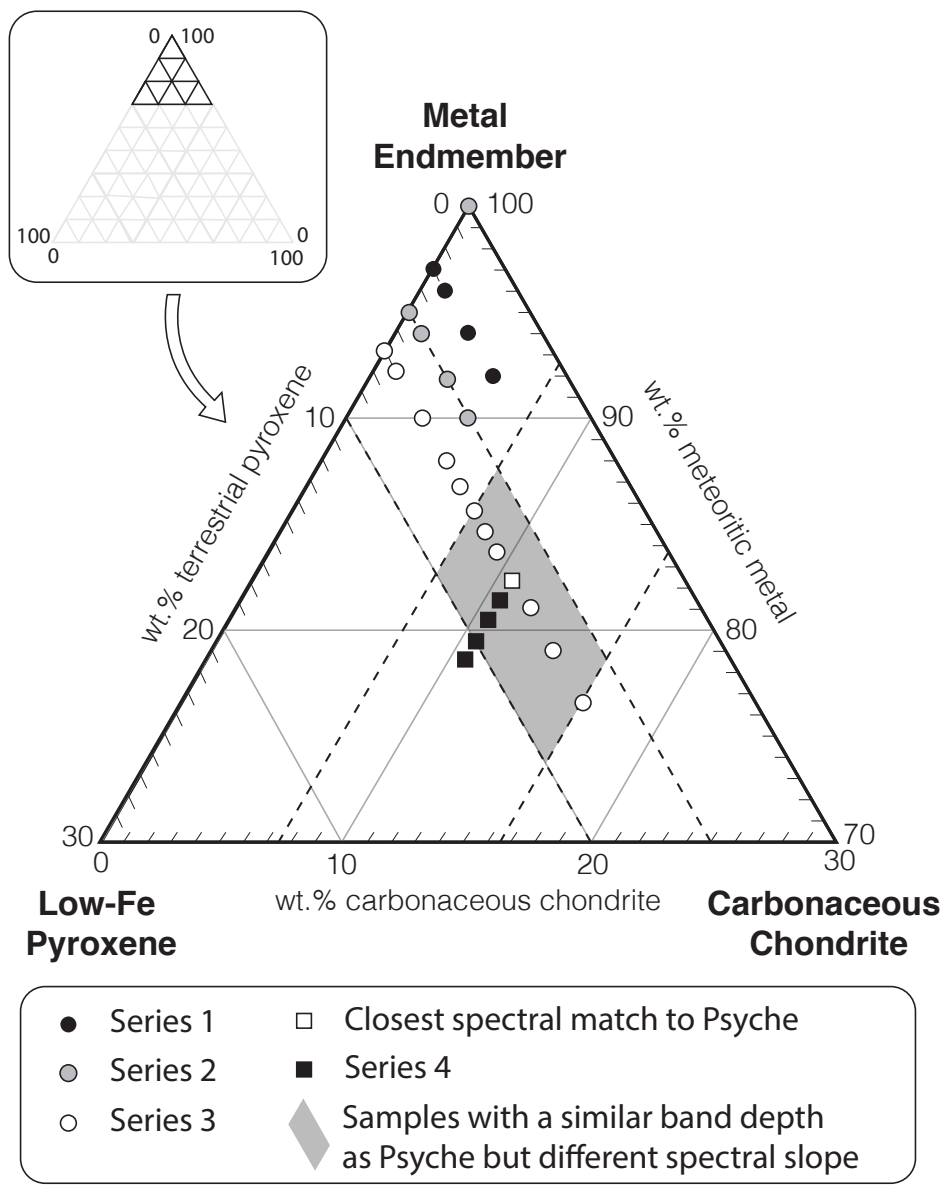}
\caption{Selected portion of a ternary diagram plotting the metal endmember, low-Fe pyroxene, and carbonaceous chondrite content of laboratory mixtures listed in Table \ref{tab:mixes}. A subplot of the full diagram is shown in the top left, highlighting how all of our mixtures were dominated by metal in composition. Series 1, 2, and 3 depict mixtures with increasing carbonaceous chondrite, decreasing metal, and fixed low-Fe pyroxene content. Series 4 depicts mixtures with increasing low-Fe pyroxene, decreasing metal, and fixed carbonaceous chondrite content. BMix 12, our best spectral match to Psyche, is shown as the white square part of both Series 3 and 4. The gray region within the plot illustrates the overall constraints placed on low-Fe pyroxene and carbonaceous chondrite content for our mixtures to match the 0.93 $\mu m$ band depth of Psyche. The addition of carbonaceous chondrite in Series 2 results in 0.93 $\mu m$ band depths that are too shallow, meaning that low-Fe pyroxene content must be greater than 5\% to offset these additions and re-deepen the band depths. Series 4 shows that any low-Fe pyroxene content greater than 10\% results in band depths that are too deep. Lastly, the addition of carbonaceous chondrite in Series 3 sets its lower and upper limits as 7.5\% and 16.5\%, respectively. These constraints were determined by analyzing mixture band depths within this series (see Section \ref{sec:results} for more information).}
\label{fig:ternary_plot}
\end{figure}

Three-component laboratory mixtures of metal (Gibeon/Georgetown), low-Fe pyroxene, and carbonaceous chondrite (Murchison) were methodically created and spectrally analyzed within the VNIR (0.35 – 2.5 $\mu m$) region (Figure \ref{fig:ternary_plot}). We compared the spectral slope and band depths of these mixtures to the telescopic spectrum of asteroid (16) Psyche in order to constrain the ratio of these three materials. Our first component, meteoritic metal, was not actively adjusted throughout our mixture sets. This endmember contained a featureless spectrum with a slope similar to that of Psyche. The addition of low-Fe pyroxene (Series 1-3) deepened the 0.93 $\mu m$ band depth and increased the reflectance between 0.6 and 0.9 $\mu m$. The addition of carbonaceous chondrite (Series 4) effectively “masked” the 0.93 $\mu m$ feature and reduced the reflectance between 0.7 and 1.2 $\mu m$. From our mixture sets, we find that:

\begin{enumerate}
    \item The mixture with the best spectral match to Psyche’s surface is composed of 82.5\% metal, 7\% low-Fe pyroxene, and 10.5\% carbonaceous chondrite by weight (61.1\% metal, 12.5\% low-Fe pyroxene, and 26.4\% carbonaceous chondrite by volume)
    \item Low-Fe pyroxene on the surface of Psyche is constrained to within 5 – 10 wt.\%
    \item Carbonaceous chondrite on the surface of Psyche is constrained to within 8 – 17 wt.\%, confirming that the surface contains hydrated material.
\end{enumerate}

Our predicted abundance of low-Fe pyroxene on Psyche’s surface (7\%) is consistent with previous results \citep[$\sim$6\%,][]{Sanchez2017}. The presence of carbonaceous chondrite also agrees with the work of \citet{Takir2017}, who observed a 3.0 $\mu m$ absorption in the near-IR. This high abundance of hydrated material on the surface of Psyche implies a complicated formation requiring large impactors as a delivery method. Assuming our best mixture set is representative of Psyche’s bulk material content, we estimate that Psyche may have a bulk porosity of $\sim$ 35\% to match the density proposed by \citet{ElkinsTanton2020}. The most recent density estimate of $2360_{-980}^{+680}$  kg m$^{-3}$ \citep{Siltala2020} may indicate an even higher bulk porosity, potentially describing a rubble pile. While Psyche may be somewhat less metallic than first thought, our results continue to describe an unusual body requiring future ground and spacecraft observations.

\acknowledgements

This work was supported by a NASA Near-Earth Object Observations (NEOO) program grant NNXAL06G (PI: Reddy). We thank Zoe Torralba for assisting in sample preparation.

\bibliographystyle{aasjournal}

\begin{deluxetable*}{c|c|c|c|c|c}[ht!]
\tablecaption{Known Physical Characteristics of Asteroid (16) Psyche\label{tab:psyche_char}}
\tablehead{\colhead{} & \colhead{Major} & \colhead{Predicted} & \colhead{Band} & \colhead{Band} & \colhead{} \\
\colhead{Diameter} & \colhead{Components} & \colhead{Abundance} & \colhead{Center} & \colhead{Depth} & \colhead{Spectral Slope} \\
\colhead{(km)} & \colhead{} & \colhead{(wt.\%)} & \colhead{($\mu m$)} & \colhead{(\%)} & \colhead{(1.8/0.8 $\mu m$ Ratio)}}
\startdata
 & Metal & $\sim 94 \pm 1$ \tablenotemark{b} & \nodata & \nodata & \\
 274 x 231 x 176\tablenotemark{a} & Low-Fe Pyroxene & $ 6 \pm 1$ \tablenotemark{b} & $0.932\pm0.006$ \tablenotemark{b} & $1.3 \pm 0.1$ \tablenotemark{b} & 1.283 \tablenotemark{b} \\
 & Carbonaceous Chondrite & Unconstrained \tablenotemark{c} & $3.00 \pm 0.03$ \tablenotemark{c} & $3.04 \pm 0.03$ \tablenotemark{c}  & \\
\hline
\enddata
\tablenotetext{a}{\citet{Drummond2018}}
\tablenotetext{b}{\citet{Sanchez2017}}
\tablenotetext{c}{\citet{Takir2017}}
\end{deluxetable*}

\begin{deluxetable*}{|c|c|c|c|}[ht!]
\tablecaption{Characteristics of Endmembers Used in Our Laboratory Experiments\label{tab:endmembers}}
\tablehead{\colhead{Endmember} & \colhead{Source} & \colhead{Type} & \colhead{(Grain Size $\mu m$)}} 
\startdata
Georgetown (Iron Meteorite) & Georgetown (Australia) & IAB Complex & $< 45$ \\
Gibeon (Iron Meteorite) & Gibeon (Namibia) & Iron IVA & $106 - 212$\\
Low-Fe Pyroxene & Mirabel Springs, California & Terrestrial Pyroxene & $< 45$ \\
Carbonaceous Chondrite & Murchison (Australia) & CM2 & $< 45$\\         
\hline
\enddata
\end{deluxetable*}

\begin{deluxetable*}{c c c c c c c}[ht!]
\tablecaption{Mixture Composition and Spectral Properties of Selected Samples Used This Experiment \label{tab:mixes}}
\tablehead{\colhead{} & \colhead{Metal} & \colhead{Low-Fe} & \colhead{Carbonaceous} & \colhead{$0.93 \mu m$} & \colhead{Spectral} & \colhead{Absolute Reflectance} \\
\colhead{} & \colhead{Endmember} & \colhead{Pyroxene} & \colhead{Chondrite} & \colhead{Band Depth} & \colhead{Slope} & \colhead{at 0.55 $\mu m$} \\
\colhead{} & \colhead{(wt.\%)} & \colhead{(wt.\%)} & \colhead{(wt.\%)} & \colhead{(\%)} & \colhead{(1.8/0.8 $\mu m$ Ratio)} & \colhead{(Albedo)}
}
\startdata
\textbf{Series 0} \\
\hline \hline
(16) Psyche & \nodata & \nodata & \nodata & $1.3 \pm 0.1$ & 1.283 & $0.093 \pm 0.024$ \\
\hline
\textbf{Series 1} \\
\hline \hline
PMix 2 & \textbf{97} & \textbf{3} & \nodata & $0.34\pm0.21$ & 1.265 & 0.071 \\
& $(97.00 \pm 0.097)$ & $(3.00 \pm 0.003)$ & & & & \\
\hline
PMix 4 & \textbf{94} & \textbf{3} & \textbf{3} & $0.26\pm0.15$ & 1.275 & 0.066 \\
 & $(94.17 \pm 0.094)$ & $(2.91 \pm 0.003)$ & $(2.91 \pm 0.003)$ & & \\
\hline
\textbf{Series 2} \\
\hline \hline
FMix 0 & \textbf{100} & \nodata & \nodata & \nodata & 1.270 & 0.073 \\
& & & & & & \\
\hline
FMix 1 & \textbf{95} & \textbf{5} & \nodata & $1.24\pm0.15$ & 1.263 & 0.072 \\
 & $(95.00 \pm 0.095)$ & $(5.00 \pm 0.005)$ & & & \\
\hline
FMix 2 & \textbf{94} & \textbf{5} & \textbf{1} & $0.50\pm0.12$ & 1.239 & 0.071 \\
 & $(94.06 \pm 0.094)$ & $(4.95 \pm 0.005)$ & $(0.99 \pm 0.001)$ & & \\
\hline
FMix 3 & \textbf{92} & \textbf{5} & \textbf{3} & $0.46\pm0.24$ & 1.267 & 0.075 \\
 & $(92.23 \pm 0.092)$ & $(4.85 \pm 0.005)$ & $(2.91 \pm 0.003)$ & & \\
\hline
\textbf{Series 3} \\
\hline \hline
BMix 2 & \textbf{93} & \textbf{7} & \nodata & $1.74 \pm 0.25$ & 1.254 & 0.075 \\
& $(93.00 \pm 0.093)$ & $(7.00 \pm 0.007)$ & & & & \\
\hline
BMix 8 & \textbf{85.5} & \textbf{7} & \textbf{7.5} & $1.56 \pm 0.21$ & 1.265 & 0.066 \\
 & $(85.62 \pm 0.086)$ & $(6.94 \pm 0.007)$ & $(7.44 \pm 0.007)$ & & \\
\hline
BMix 12 & \textbf{82.5} & \textbf{7} & \textbf{10.5} & $1.32\pm0.22$ & 1.274 & 0.069 \\
 & $(82.51 \pm 0.083)$ & $(7.02 \pm 0.007)$ & $(10.46 \pm 0.010)$ & & \\
\hline
BMix 15 & \textbf{76.5} & \textbf{7} & \textbf{16.5} & $1.18\pm0.11$ & 1.284 & 0.069 \\
 & $(76.58 \pm 0.077)$ & $(6.93 \pm 0.007)$ & $(16.49 \pm 0.016)$ & & \\
\hline
\textbf{Series 4} \\
\hline
FMix 7 & \textbf{81.5} & \textbf{8} & \textbf{10.5} & $1.40 \pm 0.19$ & 1.295 & 0.069 \\
 & $(81.60 \pm 0.082)$ & $(7.89 \pm 0.008)$ & $(10.51 \pm 0.011)$ & & \\
\hline
FMix 8 & \textbf{80.5} & \textbf{9} & \textbf{10.5} & $1.48\pm0.15$ & 1.271 & 0.071 \\
 & $(80.62 \pm 0.081)$ & $(8.99 \pm 0.009)$ & $(10.39 \pm 0.010)$ & & \\
\hline
FMix 9 & \textbf{79.5} & \textbf{10} & \textbf{10.5} & $1.50\pm0.12$ & 1.265 & 0.074 \\
 & $(79.45 \pm 0.079)$ & $(10.01 \pm 0.010)$ & $(10.54 \pm 0.011)$ & & \\
\hline
FMix 10 & \textbf{78.5} & \textbf{11} & \textbf{10.5} & $1.86\pm0.23$ & 1.253 & 0.066 \\
 & $(78.57 \pm 0.079)$ & $(11.02 \pm 0.011)$ & $(10.42 \pm 0.010)$ & & \\
\hline
\enddata
\tablecomments{Simplified weight percentages are in bold while the more precise weight and its corresponding uncertainty are below. All spectra were acquired with an incidence angle ($i$) of 0$^\circ$ and emission angle ($e$) of 30$^\circ$. }
\end{deluxetable*}

\clearpage

\end{document}